\documentclass[aps,prl,twocolumn,superscriptaddress,showpacs]{revtex4-1}
\usepackage{graphicx}% Include figure files
\usepackage{dcolumn}% Align table columns on decimal point
\begin{document}

% Use the \preprint command to place your local institutional report
% number in the upper righthand corner of the title page in preprint mode.
% Multiple \preprint commands are allowed.
% Use the 'preprintnumbers' class option to override journal defaults
% to display numbers if necessary
%\preprint{}

%Title of paper
\title{Strain and structure driven complex magnetic ordering of a 
CoO overlayer on Ir(100)}

% repeat the \author .. \affiliation  etc. as needed
% \email, \thanks, \homepage, \altaffiliation all apply to the current
% author. Explanatory text should go in the []'s, actual e-mail
% address or url should go in the {}'s for \email and \homepage.
% Please use the appropriate macro foreach each type of information

% \affiliation command applies to all authors since the last
% \affiliation command. The \affiliation command should follow the
% other information
% \affiliation can be followed by \email, \homepage, \thanks as well.
\author{F. Mittendorfer}
%\email[]{Your e-mail address}
%\homepage[]{Your web page}
%\thanks{}
%\altaffiliation{}
\affiliation{Inst. of Applied Physics, Vienna University of Technology, 
Gusshausstr. 25/134, 1040 Vienna, Austria}
\author{M. Weinert}
\email[]{weinert@uwm.edu}
%\homepage[]{Your web page}
%\thanks{}
%\altaffiliation{}
\affiliation{Department of Physics, University of Wisconsin-Milwaukee, 
Milwaukee, WI 53201, USA}
\author{R. Podloucky}
%\email[]{Your e-mail address}
%\homepage[]{Your web page}
%\thanks{}
%\altaffiliation{}
\affiliation{Inst. of Physical Chemistry, University of Vienna, Sensengasse 8, 1090 Vienna, Austria}
\author{J. Redinger}
%\email[]{josef.redinger@tuwien.ac.at}
%\homepage[]{Your web page}
%\thanks{}
%\altaffiliation{}
\affiliation{Inst. of Applied Physics, Vienna University of Technology, 
Gusshausstr. 25/134, 1040 Vienna, Austria}

%Collaboration name if desired (requires use of superscriptaddress
%option in \documentclass). \noaffiliation is required (may also be
%used with the \author command).
%\collaboration can be followed by \email, \homepage, \thanks as well.
%\collaboration{}
%\noaffiliation

\date{\today}

\begin{abstract}
We have investigated the magnetic ordering in the ultrathin c(10$\times$2) CoO(111)
film supported on Ir(100) on the basis of ab-initio calculations. We find a
close relationship between the local structural properties of the oxide film and 
the induced magnetic order, leading to alternating ferromagnetically and anti-ferromagnetically
ordered segments. While the local magnetic order is directly related to the geometric position
of the Co atoms, the mismatch between the CoO film and the Ir substrate leads to
a complex long-range order of the oxide.   

\end{abstract}

% insert suggested PACS numbers in braces on next line
% insert suggested keywords - APS authors don't need to do this
%\keywords{}

%\maketitle must follow title, authors, abstract, \pacs, and \keywords
\maketitle

% body of paper here - Use proper section commands
% References should be done using the \cite, \ref, and \label commands
%\section{Introduction}
% Put \label in argument of \section for cross-referencing
%\section{\label{intro}Introduction}

\begin{figure}
\includegraphics[width=0.4\textwidth]{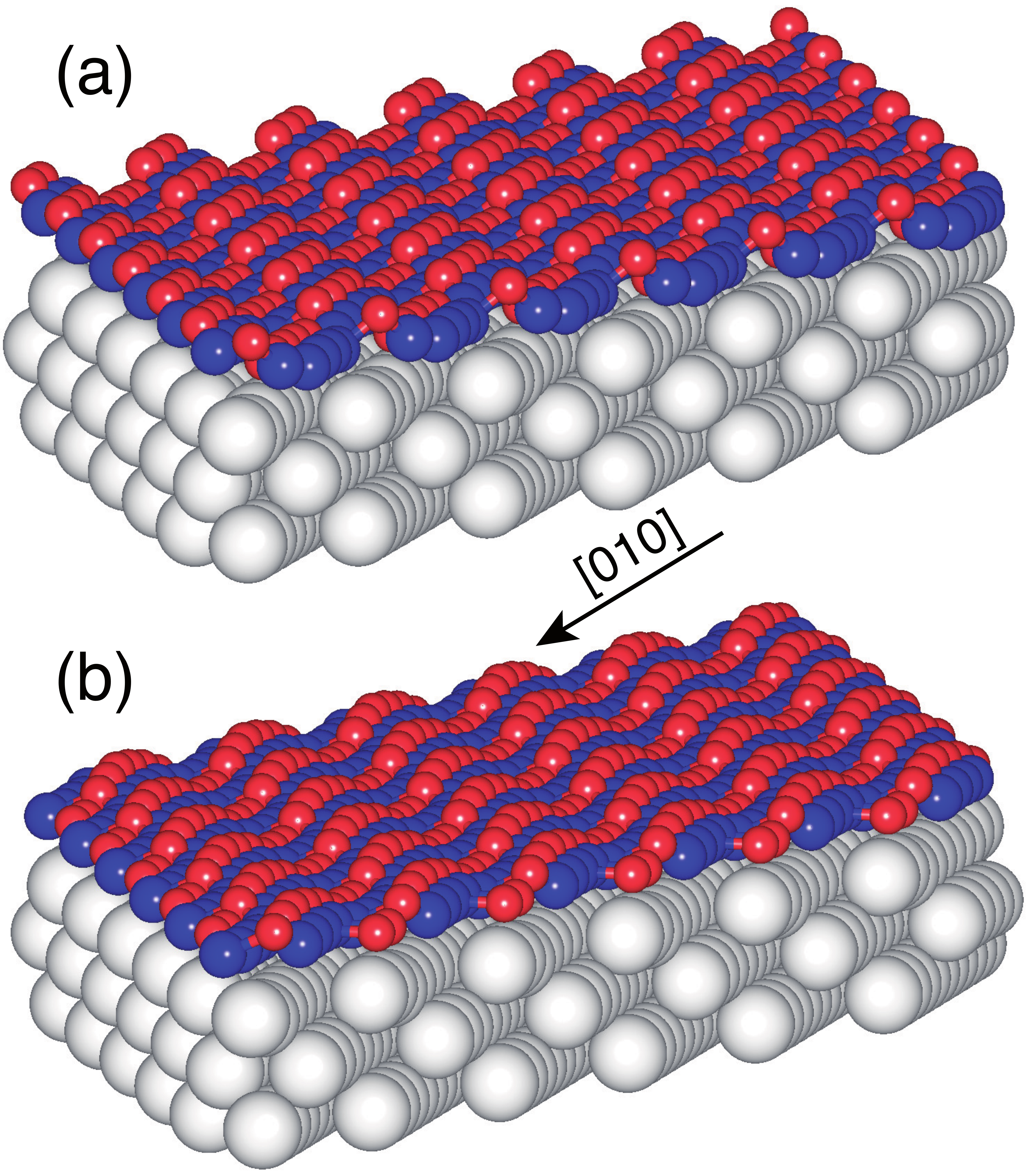}
\caption{\label{fig_aview} (Color online) Illustration of the calculated
structure for a c(10$\times$2) CoO(111) monolayer on Ir(100): red: O; blue: Co; white: Ir. 
(a) Experimentally proposed 
structure; (b) competing structure generated by shifting the CoO layer
by half the Ir-Ir distance 
along the [010] 
direction. All directions [z,x,y] are given with respect to a (1x1) Iridium cell.
See also the positional data shown in Fig. \ref{fig_struc}.}
\end{figure}

The adsorption of an ultrathin magnetic oxide film, such as CoO, on an
nonmagnetic substrate 
offers a rich and fascinating playground for studying the interplay of
the geometrical structure and the magnetic properties. 
Especially if the stable surface orientation of the oxide film
differs from the substrate, the structural 
deformations of the oxide film and the magnetic order are closely related.
This relation can be observed in the case of a CoO(111) layer supported on Ir(100), where the 
pseudo-\emph{hexagonal} atomic arrangement 
of the oxide film  has to adapt to the \emph{square} structure of the
Ir(100) substrate surface.  
Due to the  large lattice mismatch of about
9.8\% between the bulk lattice parameters of CoO(111) and Ir(100),
epitaxial growth of the CoO overlayer is very unfavorable.
For the related formation of Co$_x$O$_y$ \cite{all_1} and Mn$_x$O$_y$ \cite{cesare_1, cesare_2} films 
on Pd(100) surfaces, this lattice mismatch is compensated by the creation of Co vacancies, 
reducing the stoichiometry of the CoO film. 
Nevertheless, for the adsorption of CoO(111) on  Ir(100), the CoO film yields the 
bulk stoichiometry, but is expanded to a c(10$\times$2) overlayer,  where 9 formula units of CoO are
supported on 10 square unit cells of Ir(100) (Fig \ref{fig_aview}). 
The strain-induced relaxations in the CoO overlayer lead to sizable
structural inhomogeneities, i.e., distorted bond angles, different bond lengths,
buckling, and different local environments with respect to the substrate geometry.
Evidently,  these structural properties influence both the local and
long-range magnetic ordering of the CoO overlayer, resulting in complex
magnetic patterns.  

In the recent years, density functional theory (DFT) approaches have evolved as  powerful
tools to complement the experimental efforts to resolve the atomic structure of
supported surface oxides \cite{kli08,ser08,mit10}, and to analyze the underlying electronic
structure and the concomitant magnetic ordering.  
In this letter, we present extensive DFT calculations for two different
CoO c(10$\times$2) overlayer
structures supported on Ir(100). In addition, we demonstrate that the
close relationship between the 
magnetic configurations and the structural deformations is already present in simplified 
models of the system, thus illustrating the driving mechanism for the magnetic ordering.

Density functional theory (DFT) calculations were performed using the Vienna Ab-initio 
Simulation Package (VASP) \cite{vasp1,vasp2} in the projector augmented wave (PAW) framework \cite{paw}. 
The exchange correlation functional was described  by the general gradient approximation 
of Perdew-Burke-Ernzerhof (PBE) \cite{pbe}. To account for the localized d-states of the Cobalt 
atoms, the PBE+U approach of Dudarev \cite{ldaU} was used. 
%CHANGE 1
Varying U-J in the range between 0 to 4 eV,
we find that a value of U-J=1 eV leads to the 
best agreement with the calculated geometric structure of the CoO/Ir(100) system
with experiment.\cite{ebensperger01}
The same value of U-J=1 eV was found to yield a bulk volume very close to
experiment and also describes the electronic structure satisfactorily 
compared to a more sophisticated, but very costly, approach using hybrid functionals 
\cite{roedl}. 
In addition, we find that the most stable magnetic configuration does not depend on the 
choice of U-J. 
%END CHANGE 1
If not stated otherwise all results of the present work refer therefore to this
choice of U-J=1\,eV.

A repeated slab 
model was used consisting of  five Ir layers and CoO overlayers on both sides of the Ir slab and a 
separating  vacuum of 17~\AA, thus avoiding  dipolar interaction between 
the repeated slabs. A Monkhorst-Pack type k-mesh of
11$\times$11$\times$1 was used for small cells, and a 4$\times$4$\times$1 
k-mesh for the c(10$\times$2) structures. In all calculations the structural
degrees of freedom are fully relaxed.

Recently, an atomistic model has been proposed for the 
 c(10$\times$2)   CoO(111) layer on  Ir(100) surface on the basis of LEED experiments
\cite{ebensperger01}. 
In this study,  59 structural parameters had to be optimized to obtain a satisfactory fit, 
demonstrating the complexity of the system,
and calling for corroboration 
by an up-to-date DFT approach such as applied in this work.
Most of the experimentally derived structural parameters
\cite{ebensperger01}
agree with our DFT calculations    
within the experimental uncertainties, with only the average deviation of the  y [001] coordinates of the CoO layer
lying outside by 0.01 (O) and 0.02~\AA (Co)  \cite{ebenspergerdiplom}.
Fig.~\ref{fig_aview}(a) presents the result of the fully relaxed
calculations. In agreement with the experimental data, the oxide film
is highly corrugated and displays a pronounced corrugation in terms of height differences of the 
O ($\sim$1~\AA) and  of the  Co ($\sim$0.5~\AA) atoms. 

In this paper, we also present a second stable overlayer geometry,  
constructed by shifting the overlayer  by half the substrate
Ir-Ir  spacing in the [010] x-direction (Fig.~\ref{fig_aview}(b)). 
In this structure, which is of  different symmetry, the changes in the height of the 
O and Co atoms of the oxide film are less abrupt, although the
corrugations are 
similar to the unshifted case. Surprisingly, the energy difference  
between the these two  c(10$\times$2) structures favoring the shifted structure (Fig.~\ref{fig_aview}(b)) 
by  6 meV/Co atom is very small and negligible with respect to the expected accuracy. Note that a standard PBE
calculation tilts the balance in favor of the unshifted structure (Fig.~\ref{fig_aview}(a)) by 8 meV/Co atom. 
Therefore the DFT calculations predict that both phases coexist under
suitable experimental conditions. 
Experimentally, the two phases should be distinguishable by 
scanning tunneling microscopy (STM) because the simulated
images are distinctly different:
four bright spots (related to the uppermost oxygen atoms) are predicted
for the originally proposed structure \cite{ebensperger01}, whereas
for the new, registry-shifted structure five (or three) bright spots should
appear. Indeed, very recent experimental measurements seem to confirm the coexistence of both structures \cite{Troeppner_thesis}.

\begin{figure}
\includegraphics[width=0.45\textwidth]{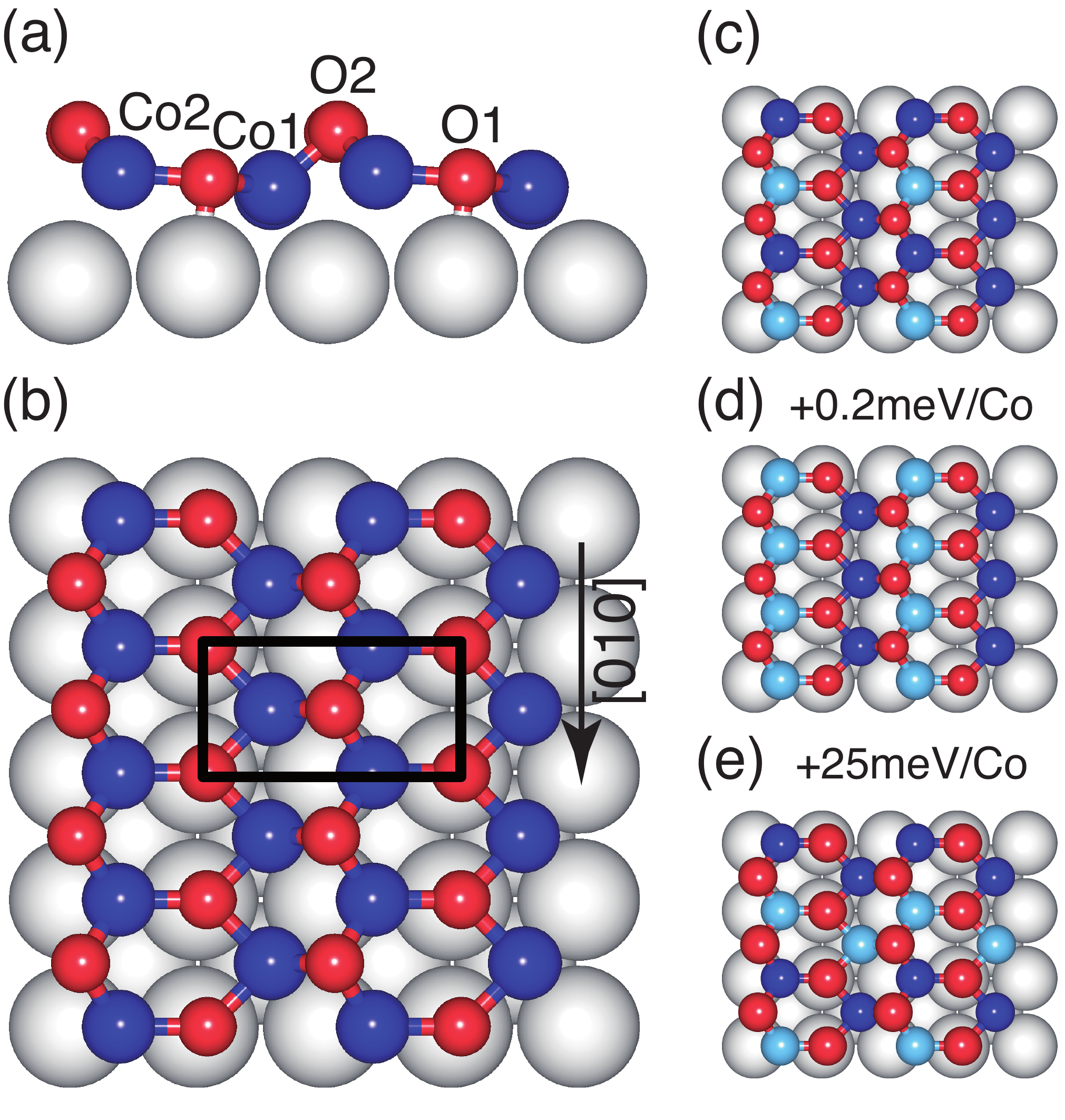}
\caption{\label{fig_SQE} (Color online) SQE model of a hexagonal
(1x1) CoO(111) monolayer on Ir(100): red: O; blue: Co; white: Ir.
(a) Side view along [010]; and (b) top view. (c)-(e) Different magnetic configurations; blue: 
Co majority, light blue: Co minority. Configurations (c) and (d) are energetically almost degenerate (see text), 
while (e) is disfavored by 25 meV/Co atom.}
\end{figure}

The basic driving forces for the structural reconstruction can be analyzed in
terms of
a simplified, quasi-epitaxial (SQE) model, where the hexagonal (111-like) CoO overlayer is accommodated 
on a square (100) substrate in a p(1$\times$2) unit cell as shown in Fig.~\ref{fig_SQE}. 
In the energetically most favorable structure, the hexagonal CoO film binds 
to the substrate via both the Co1 and Co2 atoms and the on-top (O1) oxygen atoms at the corners of the supercell.
Consequently, the Co1 and Co2 rows are pushed together by 0.53{~\AA} compared 
to the ideal positions dictated by the substrate, thus
forming Co-O1-Co bond angles between 90$^\circ$ (Co1-O1-Co1) and
135$^\circ$ (Co1-O1-Co2), that resemble
the quasi-hexagonal arrangement reminiscent of hex-BN networks. 
The second type of oxygen atoms (O2), located in bridge sites, 
is tilted out of the plane and hence does not directly contribute to the
bonding to the surface. 
The oxygen atoms display a O1-O2 buckling of 0.97~\AA, forming a
three-sided pyramid with Co-O-Co bond 
angles close to 90$^\circ$ such as found in the bulk rock-salt structure along the [111] diagonal.
Such a buckling is much less pronounced for the Co atoms, 
which are either located in hollow sites (Co1), with a  vertical
distance of 1.94{~\AA} above the substrate, or in bridge sites (Co2), at a
height of  
2.27{~\AA} (cf., Fig.~\ref{fig_SQE}(a)).  For comparison, a pure metallic Co overlayer on Ir(100) is also found to 
occupy hollow substrate sites at a height of 1.68~\AA.

The different coordination of the Co atoms is directly reflected in the magnetic ordering:
The spin moments of the Co1 atoms in the hollow sites align ferromagnetically,  
comparable to the ferromagnetic order of the pure metallic Co overlayer
(Fig.~\ref{fig_SQE}(c), dark blue atoms).
On the other hand, the magnetic order of the weaker bonded Co2 rows on the substrate bridge sites is
less well defined: 
the difference in energy between the antiferromagnetic ordering and the 
ferromagnetic ordering {\it within\/} the Co2 rows --- accompanied by an antiferromagnetic coupling to the 
Co1 rows --- is only  0.2 meV per Co atom (Figs.~\ref{fig_SQE}(c,d)).

All other reasonable (collinear) magnetic orderings are unfavorable by at least 25 meV per Co atom, 
including antiferromagnetic alignments within the Co1 rows or ferromagnetic coupling of all Co atoms.
 The induced polarization 
on the nearby Ir atoms is small since the neighboring Co atoms have partially opposite magnetic moments.
The same holds for the magnetism induced in the oxygen atoms.
Therefore we conclude that the Co atoms located on hollow sites have a strong tendency for
{\it ferromagnetic} ordering along the chains, and {\it antiferromagnetic} coupling across the lines of O atoms to the 
bridge-site Co2 chain.
Applying the  Kanamori-Goodenough-Anderson \cite{kanamori, goodenough, anderson} rules in a simple way, one would 
predict a ferromagnetic coupling of  the Co1 and Co2 chains connected by  O2 atoms via direct Co1-Co2 overlap 
(Co1-O2-Co2 bond angles close to 90$^\circ$ and Co1-Co2 distance of 2.62~\AA), 
while the coupling involving the planar O1 type atoms is less clear: The 90$^\circ$ Co2-O1-Co2  
angles and a Co2-Co2 distance of 2.74{~\AA} would again favor direct overlap ferromagnetic ordering along the Co2 rows, 
but the 135$^\circ$ Co1-O1-Co2 angles and Co1-Co2 distances of 3.56{~\AA} would certainly work against ferromagnetic 
coupling, although antiferromagnetic coupling involving the O1 atom would be strongest for 180$^\circ$ angles.
However such a simple picture for the CoO film is certainly complicated by the  interaction with the substrate, 
either by direct coupling of the Co spins via the substrate, or by changes in the oxygen orbitals
caused by bonding to the substrate that would modify the super-exchange interaction.
Thus, the magnetic ordering observed for the CoO films should be characterized as a competition between 
ferromagnetic coupling between the Co atoms by direct overlap (and also mediated by the substrate) and 
antiferromagnetic coupling across oxygen atoms in the hex-BN like structures of the layer. 

\begin{figure}
\includegraphics[width=0.45\textwidth]{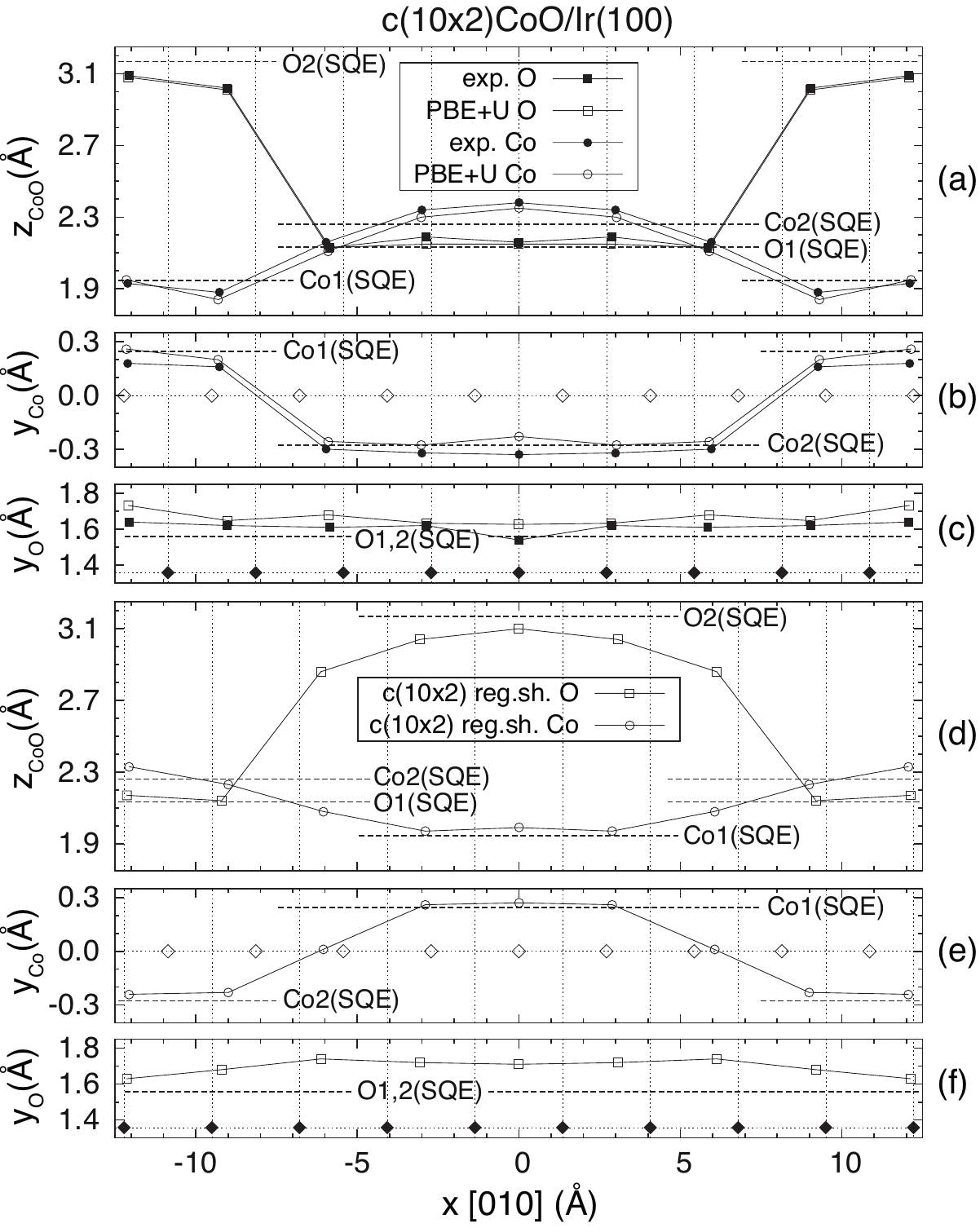}
\caption{\label{fig_struc} Structural details of a hexagonal (1x1) CoO(111) monolayer on Ir(100). (a-c): Comparison between experiment and 
PBE+U  (U-J=1\,eV) results. (d-f): CoO layer shifted by half of the substrate Ir-Ir  spacing along x [010]. Thin vertical lines show the ideal 
lateral positions of the substrate Ir atoms along  x [010], and empty
(filled) diamonds the ideal hollow (top) sites of the Ir substrate.
For comparison, the ``SQE'' labelled dashed horizontal lines denote the
atomic positions for the SQE model of  Fig.~\ref{fig_SQE}. 
The z [100] (height) values for CoO in (a) and (d) are given with respect to the mean Ir positions in the interface layer, while in 
(b-c) and (e-f) the y [001] values are given with respect to the ideal hollow sites of the Ir substrate where 
Co atoms of a metallic Co overlayer would be found. To allow for a better comparison with experiment the calculated lateral distances 
have been scaled to the experimental Ir lattice constant of 3.84~\AA (theory: 3.88~\AA).} 
\end{figure}

%CHANGE 2
However, the small energy differences are of the same order as contributions from non-collinear spin arrangements. 
Since the full c(10$\times$2) structure is too complex for a non-collinear study we
assess this possibility by performing calculations for selected non-collinear orderings between Co1 and
Co2 spins based on the SQE model (Fig. 2).  Considering Fig. 2c, as a first
step (corresponding to the collinear spin-polarized calculation)  the
collinear spin directions are fixed along [100] (out-of-plane) and then in a second step
 the FM Co1 or the AF Co2 spins are rotated by 90$^\circ$ perpendicular to [010]
(in-plane). The energy gain for these non-collinear configurations  is about 8 meV/Co.
Estimating the coupling to the substrate one finds that exchanging the spin axis of 
Co1 and Co2 favors a Co1 in-plane configuration, but 
only marginally  by 0.6meV/Co.  
Rotating the Co2 spins in-plane, while keeping the FM order (Fig. 2d), destabilizes the
configuration by 20 meV/Co with respect to the one of Fig. 2c. 
Therefore it seems likely that non-collinear spin
ordering occurs in the full c(10$\times$2) structure, but that the magnetic coupling to the substrate plays 
only a minor role here.
%END CHANGE 2

\begin{figure}
\includegraphics[width=0.45\textwidth]{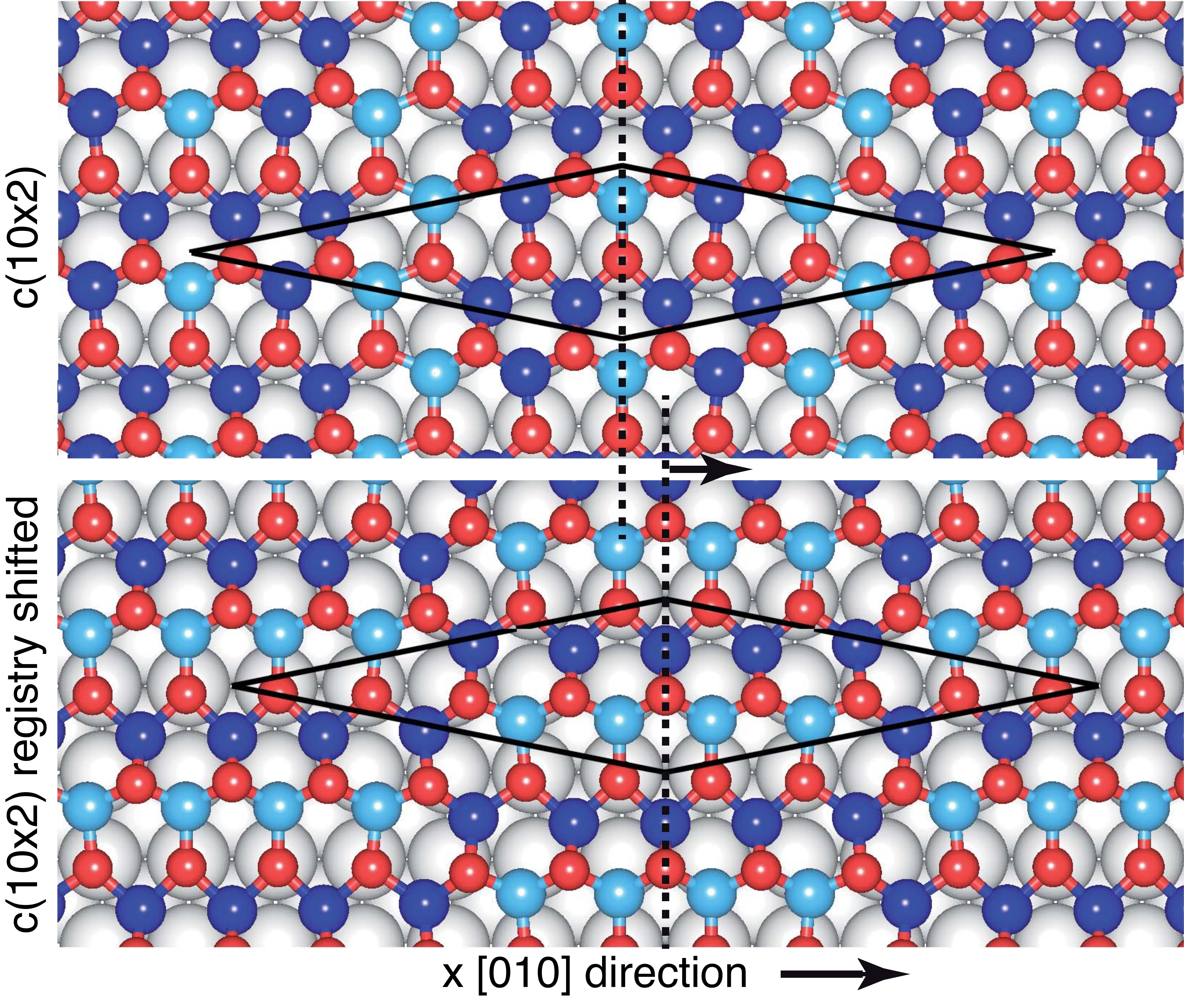}
\caption{\label{fig_c10x2_magn} (Color online) Magnetic structure of a
c(10$\times$2) CoO(111) monolayer on Ir(100): 
red: O; blue: Co majority; light blue: Co minority; white: Ir. The top panel shows the experimentally proposed
c(10$\times$2) structure\cite{ebensperger01,ebenspergerdiplom}, the
bottom panel a competing structure with a registry-shifted c(10$\times$2) CoO overlayer,
which is slightly favored by 6 meV/Co atom.}
\end{figure}

The main patterns seen in the SQE model also hold for both of the much
more complex cases of c(10$\times$2)
CoO overlayer structures on Ir(100) (see Fig.~\ref{fig_c10x2_magn}). 
Although the mismatch between the CoO film and the Ir(100) substrate leads to 
variations in the local configuration, the segments of the Co rows,
either positioned  on bridge-like or on hollow-like sites, are clearly visible
in the fully relaxed structure (Fig.\ref{fig_struc}). 
However, as the c(10$\times$2) CoO film is strained in the [010] x-direction compared to 
the SQE model, there is a continuous transition from the hollow-like to 
the bridge-like Co segments, and hence the Co atoms  are not located in the ideal
positions of the SQE model. 
Consequently, the chains in the [010] x-directions get broken,
when Co1 atoms are shifted
from their ideal hollow positions to 
bridge positions; similar arguments holds for the Co2 chains. 
Therefore, the full model does not display infinite  chains of one type of Co atoms along the [010] x-direction,
but rather alternating segments of both types arrange in a zig-zag
pattern (compare Figs.~\ref{fig_struc} and \ref{fig_c10x2_magn}). 
Still, the height and in-plane modulations for both the O and Co atoms
shown in Figs. \ref{fig_struc}(a-c)
match quite nicely the coordinates of the corresponding atoms of the
SQE model which already captures the deviations from the ideal 
positions perpendicular to the Co/O chains ([001] y-direction).
Also the agreement with the experimental LEED results \cite{ebensperger01, ebenspergerdiplom} is excellent. 

The predicted correlation between the spin alignment 
and the local geometry is also observed for the full model. 
For both CoO overlayer structures, the calculations show a particularly pronounced binding to the 
Ir substrate if the oxygen atoms are in on-top sites and 
the neighboring Co atoms are in hollow sites, forming anchor
sites around which the less binding Co2 chain fractions
are placed. This leads to the formation of two types of Co segments, a lower
segment consisting of 4 (Co1) atoms in hollow positions, and a higher 
segment of 5 (Co2) atoms. 
As already observed in the SQE model, all Co1-like atoms are
ferromagnetically aligned
(Fig.~\ref{fig_c10x2_magn}, upper panel). 
Compared with the SQE model, 
we also predict antiparallel ordering of
Co2 (bridge-like) atoms with an enhanced energy difference between 
anti- and parallel ordering within the {\it same\/} Co2 segment of 6 meV per Co atom.
%CHANGE 3
Therefore it is likely that the atoms connecting the two different chain segments 
(Co1-like and Co2-like) have a non-collinear alignment.
%END CHANGE 3
A very similar relation can be observed for the second, registry-shifted CoO
structure (Fig.~\ref{fig_c10x2_magn}, lower panel). 
Due to the shift  in the substrate lattice, there are now segments with an odd 
number of Co1 atoms (now 5) 
and an even number of Co2 atoms (now 4), and vice versa for the O atoms.
For this geometry, the spin alignments are just the ones  
expected from the SQE
model, i.e., all orderings \emph{within} the chain segments are now parallel.
%CHANGE 4
%There are  chain segments of 5 high lying O atoms, but the border atoms of
%this segments are lowered because they are significantly shifted laterally
%towards the on top substrate sites
As the non-collinear calculations for the SQE model show, a rotation of a 
\emph{ferromagnetic} Co2 spin segment is unfavorable. Hence non-collinear 
effects are expected to play a lesser role for the registry-shifted phase.  
Concerning  the chain segments of the 5 high lying O atoms, the border atoms 
of these chains are lower in height because they are significantly shifted laterally 
towards the on top  substrate sites.
%END CHANGE 4
 Overall, this leads to a surface of smoother
corrugation (cf., Figs.~\ref{fig_aview}(b) and \ref{fig_struc}(d-f))
compared to the
unshifted case. Yet, the small energy difference between the plain
registry-shifted c(10$\times$2) phases 
can lead to the coexistence of both structures, and hence
incommensurate spin ordering.

In summary, we have investigated and analyzed the complex magnetic ordering in the 
supported c(10$\times$2) CoO films on Ir(100).
The calculations predict two closely related CoO overlayer phases,
built up by Co segments consisting of either 4 or 5 atoms.
We find a close relationship between the structural relaxations and the local
magnetic ordering in the overlayer:
the segments of Co atoms above hollow-like positions of Ir(100)
clearly favor ferromagnetic coupling along the lines and antiferromagnetic coupling
between the ferromagnetic rows, while
the magnetic order is less pronounced in the segments with bridge-like atoms.
%CHANGE 5
Based on model calculations, non-collinear arrangements are expected to occur predominantly 
for the AF ordered chain segments.     
%END CHANGE 5
On the basis of these findings and their analysis, we expect a similar
relationship between structure
and magnetic ordering  for related transition-metal oxide overlayers. 
Consequently, inducing a specific deformation by the coverage by the
oxide overlayer may provide a strategy for designing magnetic properties of
complex surface oxides.

Financial support by the Austrian Science Fund (FWF): SFB FOXSI (F45) and SFB VICOM (F41), as well as by the
National Science Foundation (DMR-0706359 and DMR-1105839) is
gratefully acknowledged.
We also appreciate the computer support of the Vienna Scientific Cluster (VSC).

\end{document}